\documentclass[seceq]{ptptex}

\usepackage{graphicx}
\usepackage{wrapft}
\usepackage{fancybox}
\usepackage{subfigure}

\newcommand{\beq}{\begin{eqnarray}}
\newcommand{\eeq}{\end{eqnarray}}
\newcommand{\bfl}{\begin{flushleft}}
\newcommand{\efl}{\end{flushleft}}

\notypesetlogo                       
\title{
Comments on the ``Reply'' presented in hep-ph/0701058v1
}

\author{
Kunio \textsc{Takamatsu}
}

\inst{
KEK, High Energy Accelerator Research Organization, \\
Tsukuba, Ibaraki 305-0801, Japan\\
}

\abst{
Comments are presented on the analyses of $\pi\pi$/$K\pi$ 
scattering and production processes in 
relation with the combined fit and Adler zero 
concerning the hep-ph/0701058v1. 
}
\begin{document}
\maketitle
\section{Problems and discussions}
A reply was presented on the arXiv:hep-ph by D.V. Bugg{\cite{ref1}} 
to the article{\cite{ref2}} of myself. In this comments I shall try to 
explain further the physics view underlying our analysis, concentrating 
only on the subjects, the combined fit and Adler zero in the 
analyses of $\pi\pi$/$K\pi$ scattering and production processes near 
threshold. The cancellation mechanism in the scattering amplitudes 
is also mentioned. They may be vital points to be discussed for 
clarification of the problems still in argument. \\

\begin{flushleft}
{\it Basic physics picture underlying each analysis} \ :\\
\end{flushleft}

Before going into comments in detail, I would summarize briefly the essential 
points of our analysis on $\pi\pi$/$K\pi$ phase shift 
data and our observation 
of $\sigma$/$\kappa$ in production processes. It has long been believed 
that a production amplitude $F$ should be proportional to the scattering 
amplitude $T$ as $F=\alpha(s) T$, where $\alpha (s)$ is real and slowly 
varying function of $s$ due to the requirement from the elastic unitarity 
and the final state interaction. $\alpha (s)$ is model-dependent and 
is expressed in terms of arbitrary parameters in the conventional method. 
In the analysis of the article{\cite{ref3}} for a production process, 
the $\alpha (s)$ is expressed with the fixed zero-point
 of $T (s)$ at $s_{0}^{T}$ 
as $\alpha (s)=\alpha (s)/(s- s_{0}^T)$. Then, the huge event accumulation 
in the $\pi\pi$ production near threshold obtained at the ISR 
experiment at CERN 
was treated as backgrounds. 

Contrarily, Sigma group has analyzed the $\pi\pi$/$K\pi$ 
scattering phase shift data by the interfering amplitude method{\cite{ref4}} 
which satisfies the generalized unitarity of S-matrix. 
We have also analyzed $\pi\pi$/$K\pi$ production data 
independently from the scattering process by 
the variant mass and width method{\cite{ref5}} where all 
terms and their parameters have physical meaning,
such as coupling constant, mass and width of relevant particles. 
The base of S-matrix in our method includes unstable 
resonant particles as well as stable ones based on 
quark physics picture. 
Especially, in the case of $\pi\pi$/$K\pi$ scattering and 
production processes, the base of S-matrix includes the 
relevant unstable particle, $\sigma$/$\kappa$ as well as stable ones, 
$\pi$ and $K$. While the conventional method includes only 
stable particles. \\

\begin{flushleft}
{\it Repulsive background phase shifts} \ :\\
\end{flushleft}

The scattering phase shift data are analyzed by the 
interfering amplitude method with the introduction 
of the repulsive background phase shift, $\delta_{\rm BG}$, 
which behaves like a hard core. 
The phase shifts due to the $\sigma$ particle, $\delta_{\sigma}$ 
interferes strongly with $\delta_{\rm BG}$, resulting in the 
cancellation between them. 
The repulsive background phase shift, $\delta_{\rm BG}$ has 
the experimental base. It is recognized that $I=2$ $\pi\pi$ phase shifts, 
$\delta_{\pi\pi}^{(2)}$, where no resonance is expected, 
decrease monotonically from threshold up to $1.2 \ {\rm GeV}$, 
suggesting the existence of a hard core-like structure. 
Moreover, it has the theoretical support originating from 
the compensating $\lambda \phi^{4}$ contact interaction term 
in L$\sigma$M.{\cite{ref6}} Both $\delta_{\pi\pi}^{(2)}$ 
and $\delta_{\rm BG}$ 
obtained in the analysis show an agreement well with the 
expectation from L$\sigma$M. 

The statement in the reply {\citen{ref1}}, ``{\it The background needs 
to be parameterized empirically}'' may be from an overlooking 
of this origin. 

\section{Adler zero and the threshold suppression in production processes}
\begin{flushleft}
{\it Non appearance of threshold suppression 
in most of production processes} \ :\\
\end{flushleft}

The cancellation mechanism causes the threshold suppression 
in the $\pi\pi$/$K\pi$ scattering processes, corresponding to Adler zero. 
On the other hand in $\pi\pi$/$K\pi$ production processes though 
Adler zero exists in the processes, the threshold suppression 
does generally not occur, since the produced  $\pi\pi/K\pi$ has 
large energy transferred in most cases of production processes. 
That means Adler zero does not always result in occurrence of 
threshold suppression in a production process. 
M. Ishida et al.{\cite{ref7}} have analyzed $\pi\pi$ spectra in 
$\Upsilon$ decays, $\Upsilon(3S \to 1S)$ and $\Upsilon(2S\to1S)$. 
The results show how the threshold suppression occurs 
in the $\Upsilon(2S \to 1S)$ process and how does not in the
$\Upsilon(3S \to 1S)$, taking Adler zero in consideration. 
In the $\Upsilon(3S \to 1S)$ where the pion energy becomes enough 
larger than $m_{\pi}$ gives no zero close to the threshold. 

By the way, it is mentioned in the reply \citen{ref1} 
relating with the mechanism of the threshold suppression 
that ``{\it I see no obvious reason why the same contact term 
should appear in production process which have drastically different 
left-hand cut.}'' 
In our analysis on production processes the compensating $\lambda \phi^{4}$ 
contact interaction term plays no effective role at large energy transferred. 

Anyhow, experimentally a $\sigma$/$\kappa$ signal can mostly 
be seen explicitly in a production processes and in a certain 
case can not be, as is seen in examples mentioned above. 
This is also the reason why the production process should be 
analyzed separately from the scattering process in order to search 
for an existence of a $\sigma$/$\kappa$ resonance and why the 
resonance parameters should be determined separately from 
the scattering process in a phenomenological analysis. \\

\begin{flushleft}
{\it Combined fit and Adler zero} \ :\\
\end{flushleft}

In order to account for the difference between the 
$\pi\pi$/$K\pi$ scattering amplitude and $\pi\pi$/$K\pi$ production 
amplitudes near threshold, the suppression factor, $(s-s_{A})$ 
has been introduced explicitly into the width of the scattering 
process by the author of the reply (p.3 of ref.~\citen{ref8} 
and p.110 of ref.~\citen{ref9}). 
When the scattering amplitude $T$ is unitarized by the $N$/$D$ 
method under the prescription, $F=\alpha(s) T$, where $\alpha(s)$ 
is a real and slowly varying function of $s$ with the correction on 
$\alpha(s)$ introducing an artificial factor, $(s-s_{A})$ as $\alpha(s)
\to \{1/(s-s_{A})\} \alpha(s)$, this prescription in the $N(s)$ 
produces the $(s-s_{A})$ factor on the imaginary part of $D(s)$. 
It is also postulated in the combined fit that this $D(s)$ function 
with the $(s-s_{A})$ factor should be used for all the production processes. 
From our view point, the combined fit leads to the properties of 
relevant resonant particles which might be distorted from the proper 
ones to be determined through the quark-gluon dynamics. 
The reason is that the produced $\pi\pi$/$K\pi$ states 
in the general production processes contain not only the re-scattered 
ones but also the ones coming directly through the $ \sigma$/$\kappa$ 
production.
\section{Concluding remarks}
The difference is discussed between our method and 
the conventional one for the description of the processes, 
referring to each underlying physics picture. 
The former takes unstable particles as well as stable ones as 
the basic fields of S-matrix, while the latter takes only 
the stable particles. These two pictures may be consistent with 
each other so far as we treat interactions among stable particles, 
but will lead to decisively different methods of analyses, 
when unstable particles concern, as is seen on the mechanism 
of the threshold suppression concerning the Adler zero, for instance. 

The author of the reply mentions on the data set used in the analyses. 
The statement of the BES collaboration at the Hadron'05 
in Rio de Janeiro has answered to the problem. 


\end{document}